\documentclass[journal]{IEEEtran}

\ifCLASSINFOpdf
\else
\fi

\usepackage{amsmath}
\usepackage{cite}
\usepackage{bbold}
\usepackage{amssymb}
\usepackage{graphicx}
\usepackage{enumerate}
\usepackage{comment,color}

\newtheorem{thm1}{\bf Theorem}

\newcommand*{\QEDB}{\hfill\ensuremath{\square}}%

\begin{document}

\title{Networks obtained by Implicit-Explicit Method: \\Discrete-time distributed median solver}

\author{Jin~Gyu~Lee,~\IEEEmembership{Member,~IEEE,}
\thanks{This work was supported by the National Research Foundation of Korea(NRF) grant funded by the Korea government(Ministry of Science and ICT) (No. 2019R1A6A3A12032482).}
\thanks{J.~G.~Lee is with Control Group, Department of Engineering, University of Cambridge, United Kingdom.
{\tt\small jgl46@cam.ac.uk}}
}

\maketitle             
		
\begin{abstract}
In the purpose of making the consensus algorithm robust to outliers, consensus on the median value has recently attracted some attention.
It has its applicability in for instance constructing a resilient distributed state estimator.
Meanwhile, most of the existing works consider continuous-time algorithms and uses high-gain and discontinuous vector fields.
This issues a problem of the need for smaller time steps and yielding chattering when discretizing by explicit method for its practical use.
Thus, in this paper, we highlight that these issues vanish when we utilize instead Implicit-Explicit Method, for a broader class of networks designed by the blended dynamics approach.
In particular, for undirected and connected graphs, we propose a discrete-time distributed median solver that does not suffer from chattering.
We also verify by simulation that it has a smaller iteration number required to arrive at a steady-state.
\end{abstract}


\section{Introduction}

Consensus problem; a design problem for a network that yields agreement on their states has attracted much attention during the decades~\cite{ren2010distributed}.
Such popularity comes from its various uses in, for instance, mobile multi-robot systems and sensor networks for the purpose of coordination and estimation respectively.
In the meantime, the consensus value obtained by such couplings were usually designed or obtained as (weighted) averages of initial values or external inputs.

Meanwhile, considering its application, a large network of cheap robots or sensors, it is hard to assume the reliability of individuals, and such large-scale distributed algorithms should be resilient to faults, outliers, and malicious attacks.
In this respect, consensus on the average is inappropriate, as the mean statistic is weak to these abnormalities.
On the contrary, what is robust to outliers, is the median statistic.

By this observation, such a problem to design a network to achieve consensus to the median of initial values or external inputs has been recently tackled~\cite{franceschelli2014finite,franceschelli2016finite,dashti2019dynamic,lee2020fully,vasiljevic2020dynamic}.
Consensus on the median is still useful in the same manner illustrated earlier.
In particular,~\cite{lee2020fully} introduces its application to distributed estimation under malicious attacks.

However, most of the existing works deal with continuous-time algorithms and uses high-gain and discontinuous vector fields.
Therefore, to adjust it to implement in a discrete-time framework, the usual explicit method yields some trouble.
In particular, for stiff dynamics (high-gain), the explicit method requires a smaller time step (that depends on the gain).
But, most importantly, the used discontinuous dynamics yields chattering, which arises from the theoretical use of Filippov solution~\cite{filippov2013differential} in continuous-time.
This means a vector field can take any value in the interval, but in the end, it takes a particular value that ensures the existence of a solution.
Since such a particular value is usually hard to find, implementing this in a discrete-time framework requires an alternative. 
Otherwise, it requires a sufficiently small time step.
One way of resolving these issues is to use the implicit method.

Thus, in this paper, we will illustrate that the networks constructed by the blended dynamics approach (approach using strong diffusive coupling), such as the distributed median solver given in~\cite{lee2020fully}, can be successfully discretized by the \emph{Implicit-Explicit Method} as in~\cite{wang2019distributed}.
In particular, the obtained network does not suffer from an excess of parameters to tune; it does not have to choose an appropriate time step at each time we choose the coupling gain, unlike the explicit method.
We will concentrate on the network introduced in~\cite{lee2020fully}, as this contains discontinuity only in the individual vector field, hence makes the application of the Implicit-Explicit Method easier.
But, this is also to illustrate that this method applies well to the class of networks designed by the well-developed blended dynamics approach~\cite{lee2020tool}.

One exception in the previous works that has introduced discrete-time algorithm is~\cite{vasiljevic2020dynamic}.
However, they have another layer of network to perform the task.
The network proposed in this paper has a smaller dimension compared to this network but instead achieves only approximate consensus.

This paper is organized as follows.
In Section~\ref{sec:blen}, we briefly introduce the blended dynamics approach and illustrate how such a class of networks can be discretized by the Implicit-Explicit Method.
Then, in Section~\ref{sec:med}, we propose our discrete-time distributed median solver following the given outline and then prove its convergence.
Section~\ref{sec:sim} verifies by simulation, its ability to remove chattering, and then we conclude in Section~\ref{sec:conc}.

\section{Discretization of networks under strong diffusive coupling}\label{sec:blen}

Recently developed blended dynamics approach~\cite{lee2020tool} is based on the observation that a strong diffusive coupling makes heterogeneous multi-agent systems behave like a single dynamical system which has its vector field as the average of all the individual vector fields in the network~\cite{kim2016robustness,panteley2017synchronization}.
In particular, consider a network given as
$$\dot{x}_i = f_i(x_i) + k \sum_{j \in \mathcal{N}_i}\alpha_{ij}(x_j - x_i), \quad i \in \mathcal{N},$$
where $\mathcal{N} := \{1, \dots, N\}$ is the set of agent indices with the number of agents $N$, and $\mathcal{N}_i$ is a subset of $\mathcal{N}$ whose elements are the indices of the agents that send the information to agent $i$.
Here, the coefficient $\alpha_{ij}$ is the $ij$th element of the adjacency matrix that represents the interconnection graph, and we also assume hereafter that the graph is undirected and connected.
Then, as the coupling gain $k$ approaches infinity, the network achieves arbitrary precision approximate synchronization and its synchronized behavior can be characterized by the single dynamics 
$$\dot{\hat{x}} = \frac{1}{N}\sum_{i=1}^N f_i(\hat{x})$$
which is called blended dynamics, under the only assumption that the blended dynamics is stable, e.g., it has contraction property or has an asymptotically stable limit cycle.
The entire theory is based on the singular perturbation argument, and it has wide applicability in network design such as distributed optimization, distributed estimation, and formation control.
We refer to~\cite{lee2021design} for an exhaustive review on the topic.

Now, to avoid the problem of using the explicit method in discretizing stiff dynamics, which in this case arises as to the use of smaller time steps for increasing coupling gain~$k$ (hence making it harder to perform decentralized design),\footnote{In particular, even to ensure stability in the obtained network by the explicit method, the time step has to be well-selected to be sufficiently small.
This problem does not arise in the following approach, hence leaves the coupling gain $k$ to be the only global parameter that we have to tune.} we propose in this section the following, which makes use of the Implicit-Explicit Method.
In particular, we discretize in the manner given as
\begin{align*}
x_i[n+1] - x_i[n] &= f_i(x_i[n+1]) \\
&\quad + k\sum_{j \in \mathcal{N}_i}\alpha_{ij}(x_j[n] - x_i[n+1])
\end{align*}
or equivalently as
\begin{align*}
&x_i[n+1] + k\sum_{j \in \mathcal{N}_i}\alpha_{ij} x_i[n+1] - f_i(x_i[n+1]) \\
&\quad\quad\quad\quad\quad\quad\quad\quad\quad\quad\quad\quad\quad = x_i[n] + k\sum_{j \in \mathcal{N}_i} \alpha_{ij}x_j[n].
\end{align*}
Now, this yields the network of form
\begin{align*}
x_i[n+1] &= F_i^k\left(x_i[n] + k\sum_{j \in \mathcal{N}_i}\alpha_{ij} x_j[n]\right), \quad i \in \mathcal{N}
\end{align*}
where $F_i^k(\cdot)$ is the inverse of $(1 + kd_i)x - f_i(x)$, where $d_i = \sum_{j \in \mathcal{N}_i}\alpha_{ij}$.
By the implicit function theorem, such function $F_i^k$ is well-defined semi-globally for sufficiently large $k$.
By its definition, it satisfies the following identity.
\begin{align*}
F_i^k((1 + kd_i)x - f_i(x)) &\equiv x
\end{align*}

The utility of this approach will be found in the particular example of distributed median solver, in Section~\ref{sec:med}.

\section{Discrete-time distributed median solver}\label{sec:med}

The continuous-time distributed median solver that we want to discretize in the manner illustrated in Section~\ref{sec:blen} is motivated by the blended dynamics approach and is given in~\cite{lee2020fully} as
\begin{align}\label{eq:ct_dms}
\dot{x}_i = \text{sgn}(o_i - x_i) + k\sum_{j \in \mathcal{N}_i}\alpha_{ij}(x_j - x_i)
\end{align}
where by increasing $k$ sufficiently large, the network gets approximately synchronized to the median value of a collection $\mathcal{O}$ of real numbers $o_i$, $i = 1, \dots, N$.
The function $\text{sgn}: \mathbb{R} \to \mathbb{R}$ denotes the signum function defined as $\text{sgn}(s) = s/|s|$ for non-zero $s$, and $\text{sgn}(s) = 0$ for $s = 0$.

In this paper, the median is defined as a real number that belongs to the set
$$\mathcal{M}_\mathcal{O} := \begin{cases} \{o_{(N+1)/2}^s\}, &\mbox{ if $N$ is odd} \\ [o_{N/2}^s, o_{N/2+1}^s], &\mbox{ if $N$ is even}\end{cases} $$
where $o_i^s$'s are the elements of $\mathcal{O}$ with its index being sorted (rearranged) such that $o_1^s \le o_2^s \le \cdots \le o_N^s$.
With the help of this relaxed definition of the median, finding the median of $\mathcal{O}$ becomes solving a simple optimization problem
$$\text{minimize}_x \sum_{i =1}^N |o_i - x|.$$
Then, the gradient descent algorithm given by
$$\dot{\hat{x}} = \sum_{i=1}^N \text{sgn}(o_i - \hat{x})$$
solves the minimization problem; $\lim_{t \to \infty} \|\hat{x}(t)\|_{\mathcal{M}_\mathcal{O}} =~0$.\footnote{For a set $\Xi$, $\| x \|_{\Xi}$ denotes the distance between the vector $x$ and $\Xi$, i.e., $\| x \|_{\Xi} := \inf_{y \in \Xi} \| x - y\|$.}
This motivates the network~\eqref{eq:ct_dms} according to the blended dynamics approach introduced in Section~\ref{sec:blen}.

In this manner, we discretize the network~\eqref{eq:ct_dms} accordingly as
$$x_i[n+1] = S_i^k\left(x_i[n]+k\sum_{j \in \mathcal{N}_i}\alpha_{ij}x_j[n]\right),$$
where $S_i^k(\cdot)$ is a left inverse of $(1 + kd_i)x - \text{sgn}(o_i - x)$.
Since 
\begin{align*}
(1 + kd_i)x - \text{sgn}(o_i - x) &\!=\! \begin{cases} (1 + kd_i)x - 1, &\!\mbox{ if } x < o_i, \\ (1 + kd_i)x, &\!\mbox{ if } x = o_i, \\ (1 + kd_i)x + 1, &\!\mbox{ if } x > o_i, \end{cases}
\end{align*}
we obtain
\begin{align*}
S_i^k(x) &= \begin{cases} \frac{x + 1}{1 + kd_i}, &\mbox{ if } x < (1 + kd_i)o_i - 1,  \\ \frac{x - 1}{1 + kd_i}, &\mbox{ if } x > (1 + kd_i)o_i + 1, \\\frac{x}{1 + kd_i}, &\mbox{ otherwise} .\end{cases}
\end{align*}
In particular, the network is simply
\begin{align}\label{eq:dt_dms}
\begin{split}
x_i[n+1] &= \frac{1}{1 + kd_i}\left[1 + x_i[n] + k\sum_{j \in \mathcal{N}_i}\alpha_{ij}x_j[n]\right] \\
&\text{when } x_i[n] + k\sum_{j \in\mathcal{N}_i}\alpha_{ij}x_j[n] < (1 + kd_i)o_i - 1 \\
&= \frac{1}{1 + kd_i}\left[-1 + x_i[n] + k\sum_{j \in \mathcal{N}_i}\alpha_{ij}x_j[n]\right] \\
&\text{when } x_i[n] + k\sum_{j \in\mathcal{N}_i}\alpha_{ij}x_j[n] > (1 + kd_i)o_i + 1 \\
&= \frac{1}{1 + kd_i}\left[x_i[n] + k\sum_{j \in \mathcal{N}_i}\alpha_{ij}x_j[n]\right] \\
&\text{otherwise}
\end{split}
\end{align}
for $i \in \mathcal{N}$.
We have the following convergence result.

\begin{thm1}\label{thm:main}
Under the assumption that the communication graph induced by the adjacency element $\alpha_{ij}$ is undirected and connected, for any $\epsilon > 0$, there exists $k^* > 0$ such that, for each $k > k^*$ and initial condition $x_i[0] \in \mathbb{R}$, $i \in \mathbb{R}$, the solution to~\eqref{eq:dt_dms} exists for all $n \in \mathbb{N}$, and satisfies 
$$\limsup_{n \to \infty} \left\|x_i[n]\right\|_{\mathcal{M}_\mathcal{O}} \le \epsilon$$
for all $i \in \mathcal{N}$.
\QEDB
\end{thm1}

\begin{IEEEproof}
Note first that the network~\eqref{eq:dt_dms} can be written as
$$X[n+1] = \mathcal{B}_kX[n] + S[n]$$
where $X[n] = [x_1[n] \cdots x_N[n]]^T$, $S[n] = [s_1[n] \cdots s_N[n]]^T$, $(1+kd_i)s_i[n] \in \{-1, 0, 1\}$, and $\mathcal{B}_k$ is a stochastic matrix.
Note also that 
$$w_k := \begin{bmatrix}(1+kd_1)/\sum_{i=1}^N(1 + kd_i) \\ \vdots \\ (1+kd_N)/\sum_{i=1}^N(1 + kd_i)\end{bmatrix}$$
is the left eigenvector of $\mathcal{B}_k$ associated with the unique eigenvalue $1$.
The uniqueness comes from the connectivity of the network.
Therefore, we obtain
$$w_k^TX[n+1] = w_k^TX[n] + \frac{\sum_{i=1}^N (1 + kd_i)s_i[n]}{\sum_{i=1}^N (1 + kd_i)}$$
and
$$\left\|\mathcal{B}_k^n - 1_N w_k^T\right\| < C_kq_k^n$$
with some $C_k > 0$ and $q_k \in (0, 1)$~\cite{nedic2009distributed,nedic2014lyapunov}.
In particular, $\lim_{k \to \infty} C_k = \sqrt{\max_i d_i/\min_i d_i} =: C_\infty < \infty$ and $\lim_{k \to \infty} q_k =: q_\infty \in (0, 1)$.
See Appendix~\ref{app:ratio} for its illustration.

Now, since
\begin{align*}
X[n] &= \mathcal{B}_k^nX[0] + \sum_{i=1}^n \mathcal{B}_k^{n-i}S[i-1]
\end{align*}
we can conclude that
\begin{align*}
(I_N - 1_Nw_k^T)X[n] &= (\mathcal{B}_k^n - 1_Nw_k^T)X[0] \\
&\quad + \sum_{i=1}^n (\mathcal{B}_k^{n-i} - 1_Nw_k^T)S[i-1].
\end{align*}
Therefore,
\begin{align*}
&\left\|(I_n - 1_Nw_k^T)X[n]\right\| \\
&\le \left\|\mathcal{B}_k^n - 1_Nw_k^T\right\|\!\left\|X[0]\right\| \!+\! \sum_{i=1}^N \left\|\mathcal{B}_k^{n-i} \!-\! 1_Nw_k^T\right\|\!\left\|S[i-1]\right\| \\
&\le C_k q_k^n\left\|X[0]\right\| + \sum_{i=1}^n C_k q_k^{n-i}\frac{\sqrt{N}}{1 + k\min_i d_i} \\
&\le C_k q_k^n\left\|X[0]\right\| + \frac{C_k}{1 - q_k}\frac{\sqrt{N}}{1 + k\min_i d_i}.
\end{align*}

This implies that for any $\epsilon > 0$, there exists $k^*$ such that for each $k \ge k^*$ and initial condition $X[0]$, there exists $n^* \in \mathbb{N}$ such that
\begin{align}\label{eq:app}
\left\|(I_n - 1_Nw_k^T)X[n]\right\| \le \frac{\epsilon}{3}
\end{align}
for all $n \ge n^*$.
Note that $C_k/(1-q_k) > 1$, and thus, we have
$$\frac{\epsilon}{3} > \frac{1}{1+kd_i}, \quad \forall i \in \mathcal{N}$$
for such $k$.

Now, since we have proved arbitrary precision approximate synchronization, let us recall how the averaged variable $w_k^TX[n]$ behaves;
$$w_k^TX[n+1] = w_k^TX[n] + \frac{\sum_{i=1}^N \hat{s}_i[n]}{\sum_{i=1}^N(1 + kd_i)}$$
where $\hat{s}_i[n] := (1 + kd_i)s_i[n]\in \{-1, 0, 1\}$.
This implies that if the overall balance $\sum_{i=1}^N \hat{s}_i[n]$ is positive, then the averaged value increases, while if the balance is negative, then the value decreases.
On the other hand, by its construction $\hat{s}_i[n] = -1$ if and only if (see~\eqref{eq:dt_dms})
$$x_i[n] + k\sum_{j \in \mathcal{N}_i}\alpha_{ij} x_j[n] > (1 + kd_i) o_i + 1,$$
and thus, for $n \ge n^*$, if
$$w_k^TX[n] \ge o_i  + 2\frac{\epsilon}{3},$$
then by~\eqref{eq:app}, we have
\begin{align*}
x_i[n] + k\sum_{j \in \mathcal{N}_i}\alpha_{ij} x_j[n] &\ge (1 + kd_i)\left(w_k^TX[n] - \frac{\epsilon}{3}\right) \\
&\ge (1 + kd_i) \left(o_i + \frac{\epsilon}{3}\right) \\
&> (1 + kd_i)o_i + 1
\end{align*}
hence $\hat{s}_i[n] = -1$.
Similarly, if
$$w_k^TX[n] \le o_i - 2\frac{\epsilon}{3},$$
then we have $\hat{s}_i[n] = 1$.
This finally implies that if
$$w_k^TX[n] \ge \max \mathcal{M}_\mathcal{O} + 2\frac{\epsilon}{3},$$
then we have $\sum_{i=1}^N\hat{s}_i[n] \ge 1$, hence the averaged value increases for at least $1/\sum_{i=1}^N(1+kd_i)$ amount, and similarly, if
$$w_k^TX[n]\le \min \mathcal{M}_\mathcal{O} - 2\frac{\epsilon}{3},$$
then we have $\sum_{i=1}^N\hat{s}_i[n]\le -1$, hence the averaged value decreases accordingly.
Therefore, we can conclude that
$$\limsup_{n \to \infty} \left\|w_k^TX[n]\right\|_{\mathcal{M}_\mathcal{O}} \le 2\frac{\epsilon}{3}$$
which concludes the proof with the help of~\eqref{eq:app}.
\end{IEEEproof}

Now, in the next section, we observe in the simulation result, a dramatic removal of the chattering phenomenon, compared to the network obtained by the explicit method. 

\section{Simulation}\label{sec:sim}

\begin{figure}[h]
\begin{center}
\includegraphics[width=\columnwidth]{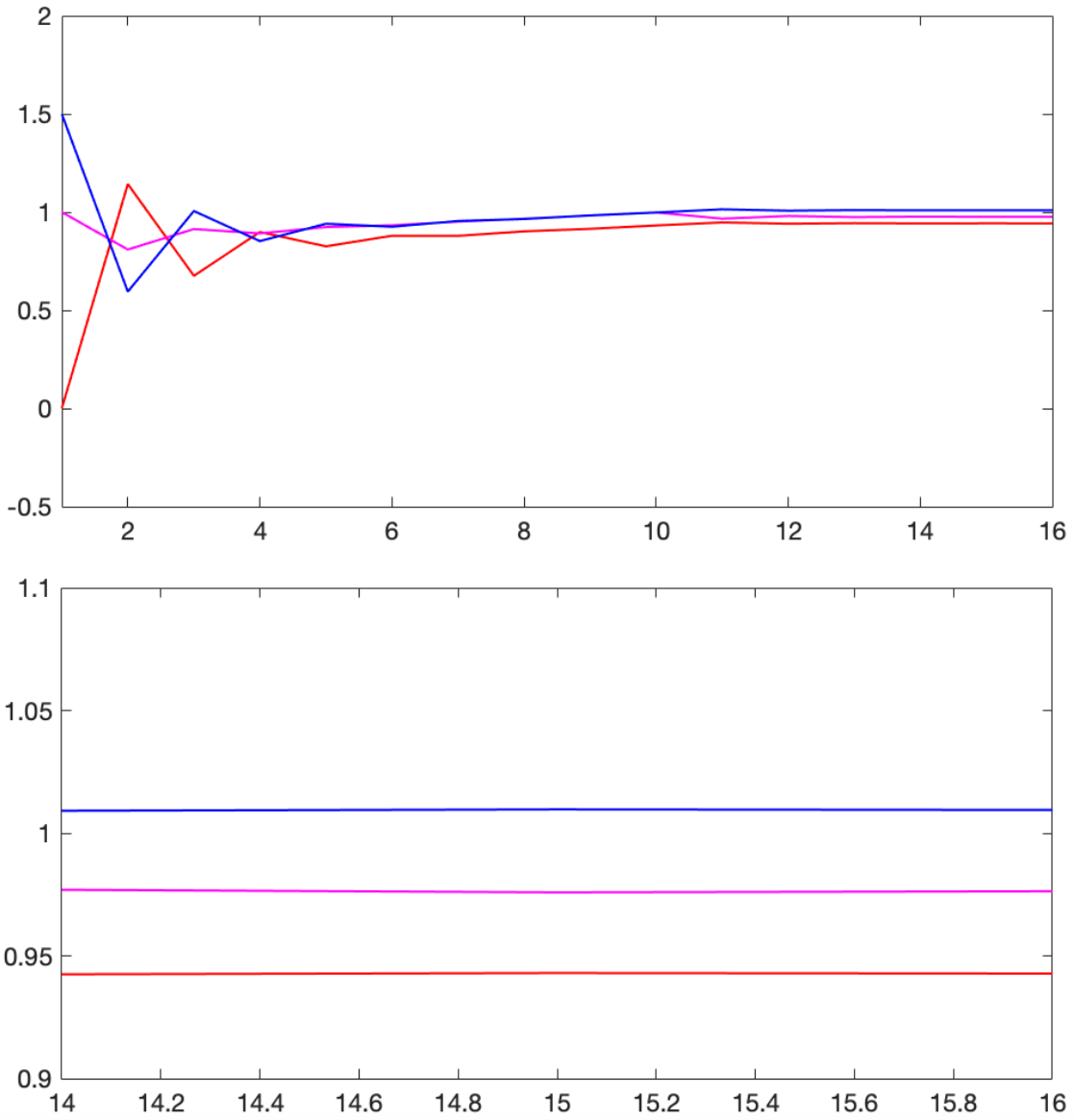}
\caption{Simulation result of the network~\eqref{eq:dt_dms} with initial conditions $x_1[0] = 0$, $x_2[0] =1$, and $x_3[0] = 1.5$.}
\label{fig:1}
\end{center}
\end{figure}

To compare the chattering phenomenon in the network, we consider a simple network consisting of three agents, where $o_1 = 0$, $o_2 = 1$, and $o_3 = 100$.
The communication graph is complete and unitary; $\alpha_{ij} = 1$ for all $i \neq j$.
The simulation result of the network~\eqref{eq:dt_dms} with $k = 10$ is given in Figure~\ref{fig:1}.

\begin{figure}[h]
\begin{center}
\includegraphics[width=\columnwidth]{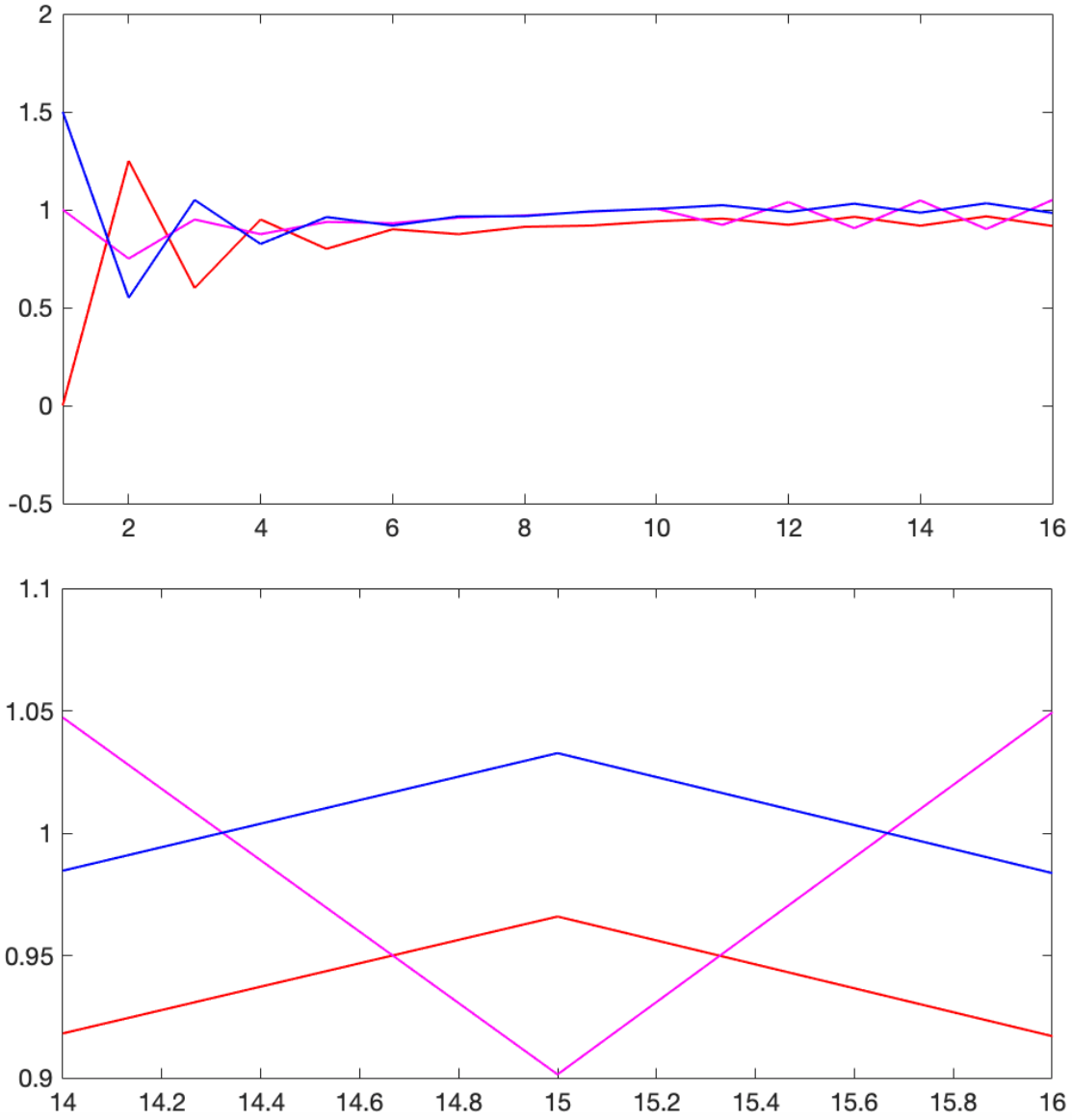}
\caption{Simulation result of the network~\eqref{eq:dt_dms_E} with $T_s = 0.05$ and initial conditions $x_1[0] = 0$, $x_2[0] =1$, and $x_3[0] = 1.5$.}
\label{fig:2}
\end{center}
\end{figure}

On the other hand, if we simulate a network obtained by discretizing~\eqref{eq:ct_dms} with the explicit method, given as
\begin{align}\label{eq:dt_dms_E}
\begin{split}
x_i[n+1] &= x_i[n] + T_s\text{sgn}(o_i - x_i[n]) \\
&\quad + k T_s \sum_{j \in \mathcal{N}_i}\alpha_{ij}(x_j[n] - x_i[n])
\end{split}
\end{align}
for sufficiently small time step $T_s > 0$, then with $T_s = 0.05$, we get the trajectories illustrated in Figure~\ref{fig:2}.
Here, by slightly increasing the time step to $T_s = 0.07$, we get unstable trajectories.

\begin{figure}[h]
\begin{center}
\includegraphics[width=\columnwidth]{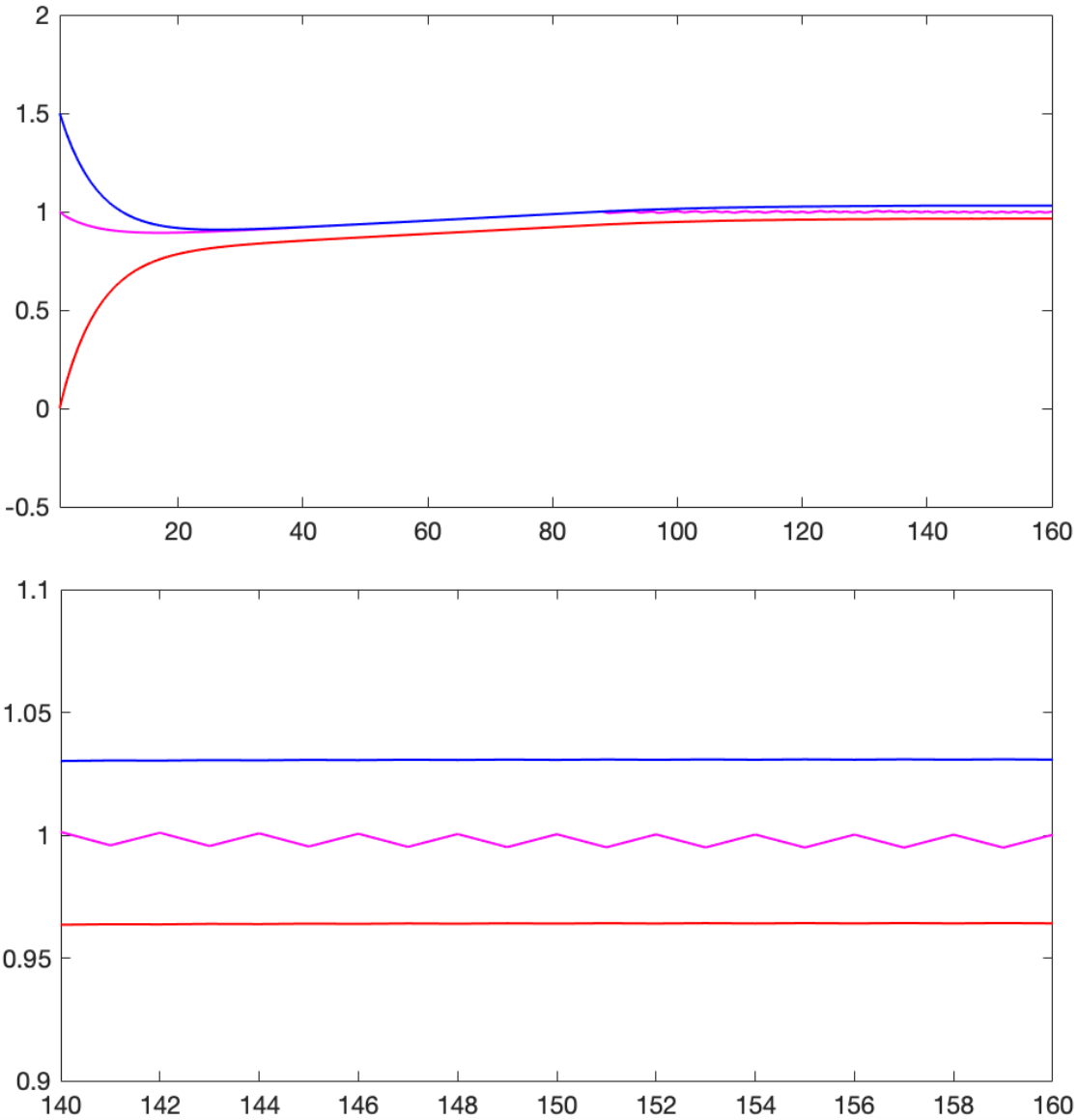}
\caption{Simulation result of the network~\eqref{eq:dt_dms_E} with $T_s = 0.005$ and initial conditions $x_1[0] = 0$, $x_2[0] =1$, and $x_3[0] = 1.5$.}
\label{fig:3}
\end{center}
\end{figure}

Note that we observe not only a bigger steady-state error, but also the chattering phenomenon.
To recover the accuracy in the steady-state limit, we should employ $T_s = 0.005$, which results in the simulation result given in Figure~\ref{fig:3}, but it requires a larger number of iteration.

\section{Conclusion}\label{sec:conc}
By studying for a particular example of distributed median solver, we have seen the utility of the Implicit-Explicit Method, in discretizing a network constructed by the blended dynamics approach, which for the explicit method, by its stiffness in the dynamics, suffers from a problem like the need for a smaller time step, which depends on the coupling gain.
Moreover, the method has proven its ability to remove the chattering phenomenon when discretizing a system having discontinuity in its vector field.
The future consideration will be on the general conclusion of the use of the Implicit-Explicit Method on such class of networks and also on the analytical verification of its advantages compared with other methods.

\bibliographystyle{IEEEtran}
\bibliography{Reference}

\appendix

\section{Illustration of $C_k$ and $q_k$ in the proof of Theorem~\ref{thm:main}}\label{app:ratio}

First define the Laplacian matrix $\mathcal{L} = [l_{ij}] \in \mathbb{R}^{N \times N}$ of a graph as $\mathcal{L} := \mathcal{D} - \mathcal{A}$, where $\mathcal{A} = [\alpha_{ij}]$ is the adjacency matrix of the graph and $\mathcal{D}$ is the diagonal matrix whose diagonal entries are $d_i$, $i \in \mathcal{N}$.
By its construction, it contains at least one eigenvalue of zero, whose corresponding eigenvector is $1_N := [1,\dots,1]^T \in \mathbb{R}^N$, and all the other eigenvalues have nonnegative real parts.
For undirected graphs, the zero eigenvalue is simple if and only if the corresponding graph is connected.
Moreover, $I_N - \mathcal{D}^{-1}\mathcal{L}$ has its eigenvalues contained inside the unit circle and the eigenvalue with magnitude one becomes unique if and only if the graph is connected.

Then, we have the representation
$$I_N - \mathcal{B}_k = \text{diag}\left(\frac{k}{1+kd_1}, \dots, \frac{k}{1+kd_N}\right)\mathcal{L} =: \mathcal{D}_k\mathcal{L},$$
and thus,
$$I_N - \sqrt{\mathcal{D}_k}^{-1}\mathcal{B}_k \sqrt{\mathcal{D}_k} = \sqrt{\mathcal{D}_k}\mathcal{L}\sqrt{\mathcal{D}_k} =: \mathcal{L}_k.$$
Since $\mathcal{L}_k$ is a symmetric positive semi-definite matrix, there exists normalized eigenvectors $v_{1, k}, \dots, v_{N, k}$ associated with eigenvalues $0 < \lambda_{2,k} \le \dots \le \lambda_{N, k}$ such that 
$$\mathcal{L}_k v_{i, k} = \lambda_{i, k}v_{i, k} \quad \text{ and } \quad v_{i, k}^T\mathcal{L}_k = \lambda_{i, k} v_{i, k}^T$$
for $i = 2, \dots, N$ and $\mathcal{L}_kv_{1, k} = 0$, $v_{1, k}^T\mathcal{L}_k = 0$.
This implies that 
$$\sqrt{\mathcal{D}_k}^{-1}\mathcal{B}_k\sqrt{\mathcal{D}_k} = \mathcal{V}_k \text{diag}(1, 1 - \lambda_{2, k}, \dots, 1 - \lambda_{N, k})\mathcal{V}_k^T$$
where $\mathcal{V}_k = [v_{1, k} \cdots v_{N, k}]$.
Therefore, by noting that $v_{1, k}$ is the normalized vector of $\sqrt{\mathcal{D}_k}^{-1}1_N$, hence
$$\sqrt{\mathcal{D}_k}\mathcal{V}_k\text{diag}(1, 0, \dots, 0)\mathcal{V}_k^T\sqrt{\mathcal{D}_k}^{-1} = 1_Nw_k^T$$
we can conclude that
\begin{align*}
&\mathcal{B}_k^n - 1_Nw_k^T= \\
&\sqrt{\mathcal{D}_k}\mathcal{V}_k\text{diag}\left(0, (1 - \lambda_{2, k})^n, \dots, (1 - \lambda_{N, k})^n\right) \!\mathcal{V}_k^T\!\sqrt{\mathcal{D}_k}^{-1}.
\end{align*}
This finally implies 
\begin{align*}
\left\|\mathcal{B}_k^n - 1_N w_k^T\right\| &\le C_k q_k^n
\end{align*}
where
\begin{align*}
C_k &:= \left\| \sqrt{\mathcal{D}_k}\right\| \left\|\sqrt{\mathcal{D}_k}^{-1}\right\| = \sqrt{\frac{1 + k\max_i d_i}{1 + k\min_i d_i}}, \\
q_k &:= \max\left\{\left| 1 - \lambda_{2, k}\right|, \left| 1 - \lambda_{N, k}\right|\right\}.
\end{align*}
Now, noting that $\mathcal{L}_k$ has the same set of eigenvalues with the matrix $\mathcal{D}_k\mathcal{L}$, by the Gershgorin circle theorem, we can certify that $\lambda_{i, k} \in[0, 2)$, hence $q_k < 1$.
In particular, we have
\begin{align*}
\lim_{k \to \infty} C_k &= \sqrt{\frac{\max_i d_i}{\min_i d_i}} \\
\lim_{k \to \infty} q_k &= \max\left\{\left|1 - \lambda_{2, \infty}\right|, \left|1 - \lambda_{N, \infty}\right|\right\}
\end{align*}
where $0 < \lambda_{2, \infty} \le \cdots \le \lambda_{N, \infty} < 2$ are the eigenvalues of
$$\text{diag}(1/d_1, \dots, 1/d_N)\mathcal{L} = \mathcal{D}_\infty \mathcal{L} = \mathcal{D}^{-1}\mathcal{L}.$$

\end{document}